\documentclass[twocolumn,showpacs,amsmath,amssymb,prl,superscriptaddress]{revtex4-1}

\usepackage{graphicx}
\usepackage{bm}
\usepackage{color,ulem}

\begin{document}

\title{Coulomb-enhanced resonance transmission of quantum SINIS junctions}
\author{N. B. Kopnin}
\affiliation{Low Temperature Laboratory,  Aalto
University, PO Box 15100, FI-00076 AALTO, Finland}
\affiliation{L. D. Landau Institute for Theoretical
Physics, 117940 Moscow, Russia}
\author{Y.  M. Galperin}
\affiliation{Department of
Physics, University of Oslo, PO Box 1048 Blindern, 0316 Oslo,
Norway}
\affiliation{A. F.  Ioffe Physico-Technical
Institute of Russian Academy of Sciences, 194021 St. Petersburg,
Russia}
\author{V. M. Vinokur}
\affiliation{Argonne National Laboratory, 9700 S.
Cass Ave., Argonne, IL 60439, USA}
\date{\today}

\begin{abstract}

Coherent charge transfer through a ballistic gated SINIS junction
is mediated by the resonant tunneling via the Andreev states.
Extra charge accommodated on the Andreev levels partially
compensates the charge induced by the gate voltage preserving the
electron wavelength and maintaining the resonance conditions in a
broad range of gate voltages. As a result, the transparency of the
junction as well as the supercurrent trough it can be
substantially increased as compared to the zero-Coulomb case.
\end{abstract}

\maketitle

According to the classical picture of the Coulomb blockade in
gated structures~\cite{blockade} the Coulomb interaction usually
suppresses electronic transport through small conductors by
introducing an additional energy barrier associated with the
charging energy. In this Letter we show that in ballistic SINIS
junctions (here S stands for a superconductor, N is a normal-metal
island, and I is an insulator), the Coulomb interaction can rather
\textit{stimulate} the supercurrent than suppress it. The
supercurrent through a junction containing a ballistic normal
island is mediated by Andreev states with energies controlled by
the effective transparency of the double-barrier structure which
is determined by the wave length of electrons. The extra charge on
the dot affects not only the energy but also the wave length of
electrons. This makes the transmission extremely sensitive both to
the gate voltage and to the charge on the Andreev states localized
in the normal island. We show that charging of Andreev states can
actually \textit{preserve} the condition of resonance tunneling:
the charge adjusts itself so as to compensate the deviation of the
chemical potential caused by the change in the gate voltage thus
increasing the transparency of the device for the supercurrent.

\paragraph{The system. --}
We consider the normal island in the form of a short single-mode
ballistic conductor (wire) connected to bulk superconducting leads
via low-transparency contacts. This corresponds to the common
experimental situation which is realized, in particular, in recent
experiments on nanowires~\cite{nanotubes1}. The length $d$ of a
conductor is much less than the superconducting coherence length,
$\xi$, i.e., $|\Delta|\ll \delta E $ where $\delta E\sim\hbar
v_F/d$ is the level spacing in the double-barrier structure, $v_F$
is the velocity of electrons in a quantum conductor, and
$|\Delta|$ is the superconducting gap in the leads. The short
conductor (sometimes called an Andreev quantum dot) is weakly
coupled to the superconducting leads and forms a low-proximity
device. This setup is similar to that considered in
Refs.~\cite{Sadovskii07} where the charging of Andreev levels was
studied in the limit of small Coulomb interaction.

To characterize the efficiency of the Coulomb interaction in the
SINIS we introduce the dimensionless parameter $\gamma=E_C/\delta
E\sim e^2d/C\hbar v_F$, where  $E_C=e^2/2C$ is the Coulomb energy
of the island, $C$ being its capacitance. The maximum value of
$\gamma$ is less than $ e^2/\hbar v_F \varkappa $, with the island
capacitance $C\sim \varkappa d\ln (d/a)$, where $a$ is the
transverse dimension of the wire and $\varkappa$ is the effective
dielectric constant in the presence of a substrate.  For a typical
value $v_F\sim 10^{8}$~cm/s the corresponding $\gamma$ is small;
thus the assumption $\gamma\ll 1$ is adequate for practical
devices. The same parameter characterizes the Luttinger-liquid
effects~\cite{Dolcini03}, which thus can be neglected. However, in
spite of the small value of $\gamma$, the Coulomb energy can still
be either larger or smaller than the superconducting gap. To
account for nonlinear back-action effects on the contact
transparency associated with charging of the Andreev level in case
of a considerable Coulomb energy $E_C\sim |\Delta|$, we adopt a
mean-field model similar to that employed in Ref.~\cite{KGV09} for
normal junctions. In this respect, our model differs from those
considering transport through quantum dots having a single {\it
fixed} electron level~\cite{qd}.

\paragraph{Model. --}
 The charging Hamiltonian has the form
\begin{equation}
\hat H_c =E_{C}(\hat N -N_0)^2 \, , \quad
\hat N =\sum _\alpha \int _V \hat \psi ^\dagger_\alpha ({\bf
r})\hat \psi _\alpha ({\bf r})\, d^3r \, , \label{Noperator}
\end{equation}
the sum being taken over the spin indices $\alpha$ and the
integration domain is the volume $V$ of the island. Here $N_0$ is
the background charge supplied by ions and by the external
circuit. The kinetic energy has the usual form,
\[
H_{\text{kin}}=\sum_\alpha\int d^3 r \, \hat \psi_\alpha ^\dagger
({\bf r}) \left[-\frac{\hbar^2\nabla^2}{2m}-E_F+V({\bf
r})\right]\hat \psi_\alpha ({\bf r}),
\]
where $V({\bf r})=I[\delta (x+d/2)+\delta(x-d/2)]$ describes the
insulating barrier between the island and the leads, $x$ is the
coordinate along the conductor.

Superconductivity in the leads is
described by the usual BCS Hamiltonian where $\Delta({\bf  r})
=|\Delta|e^{i\chi_{L,R}}$ inside the leads, i.e., for $x<-d/2$ or
$x>d/2$, respectively, while $\Delta =0$ in the normal conductor,
$-d/2<x<d/2$.

In this model, the states of the system are spin-independent. The
advanced and retarded Green functions in the real-frequency
representation can be found from the Dyson equation
\begin{equation}
\int d^3 r \check L(\mathbf{r}_1,\mathbf{r})\check
G_{\epsilon}^{R(A)}({\bf r},{\bf r}_2)=\delta({\bf r}_1-{\bf r}_2)
\end{equation}
where
\begin{eqnarray*}
&&\check L ({\bf r}_1, {\bf r}) = \check G_0^{-1}(\mathbf{r}_1)\delta({\bf r}
-{\bf r}_1)+2E_C h({\bf r}_1) \check K (\mathbf{r}_1,\mathbf{r}) \, , \nonumber \\
&&\check G_0^{-1}(\epsilon,{\bf r}) =\left( \begin{array}{cc}
-\epsilon +\hat H_0 & -\Delta \\
\Delta^* &\epsilon +\hat H_0\end{array}\right), \  \check G
=\left( \begin{array}{cc} G & F\\ -F^\dagger & \bar
G\end{array}\right) ,  \nonumber \\
\end{eqnarray*}
are matrixes in the Nambu space, and
\begin{eqnarray*}
&&\hat H_0=-\hbar^2{\nabla}^2/2m -E_F+V({\bf r})+U_Ch({\bf r}) \, ,  \nonumber \\
&&  U_C=2E_C(N-N_0), \  N=-\int_V d^3 r \, {\rm Tr}\,
\check K({\bf r},{\bf r}) \, . \nonumber
\end{eqnarray*}
The function $h({\bf r})=1$ if ${\bf r}$ belongs to the island and
$h({\bf r})=0$ otherwise. The kernel
\begin{equation}
\check K({\bf r}_1,{\bf r})=\int \frac{d\epsilon }{4\pi i}\check
G^K_{\epsilon } ({\bf r}_1,{\bf r})
\end{equation}
is determined by the Keldysh function, $ \check
G_{\epsilon}^{K}({\bf r}_1,{\bf r}_2)=\left[\check
G_{\epsilon}^{R}({\bf r}_1,{\bf r}_2)-\check G_{\epsilon}^{A}({\bf
r}_1,{\bf r}_2)\right]f_1(\epsilon) $. Here
$f_1(\epsilon)=1-2n_\epsilon$ and $n_\epsilon$ is the occupation
number. In equilibrium $n_\epsilon$ is the Fermi function, and
$f_1(\epsilon)=\tanh(\epsilon/2T)$. The energy $U_C$ describes the
variation of the charge on the island due to the change in the
number of excitations   while $N$ has a meaning of  the excitation
charge on the island.

\paragraph{Bogoliubov-de Gennes (BdG) equations. --}
Let us expand the retarded and advanced Green functions
in a set of orthogonal and normalized functions as
$$\check G_{\epsilon}^{R(A)}({\bf r}_1,{\bf r}_2)=\sum _n
\frac{\check U_n({\bf r}_1) \check U_n^\dagger ({\bf r}_2)}{E_n -
\epsilon \mp i\delta}, \ \check U_n({\bf r})=
\left( \! \! \!\begin{array}{c} \phantom{-}u_n({\bf r})
\\-v_n({\bf r})\end{array} \! \! \right)\!,$$
$\check U_n^\dagger=(u_n^*\ , v_n^*)$. The functions $u$ and $v$
satisfy the linear equations with effective potentials determined
by $ K_{uu}({\bf r},{\bf r}_1)$ and $ K_{uv}({\bf r},{\bf r}_1)$:
\begin{eqnarray}
&& \hat H_0 u_n({\bf r}) +\!\Delta({\bf r}) v_n({\bf r}) +
E_Ch({\bf
r})\!\! \int_{V}\!\! K_{uu}({\bf r},{\bf r}_1\!) u_n({\bf r}_1\!)d^3 r_1 \nonumber \\
&&\quad -E_Ch({\bf r})\int_{V} K_{uv}({\bf r},{\bf r}_1\!) v_n({\bf
r}_1\!) d^3 r_1 =E_n u_n({\bf r})\, ,
\label{inteq1}\\
&&\hat H_0 v_n({\bf r}) -\!\Delta^* ({\bf r}) u_n({\bf r}) + \!
E_Ch({\bf
r})\!\! \int_{V}\!\! K_{uu}^*({\bf r},{\bf r}_1\!) v_n({\bf r}_1\!) d^3 r_1 \nonumber \\
 &&\quad +E_Ch({\bf r}) \int_{V} K_{uv}^*({\bf r},{\bf r}_1\!) u_n({\bf r}_1\!) d^3r_1
=-E_nv_n({\bf r})\, . \label{inteq2}
\end{eqnarray}
The kernels $ K_{uu}({\bf r},{\bf r}_1)$ and $ K_{uv}({\bf r},{\bf
r}_1)$ are determined self-consistently. They can also be expanded
into the same wave functions,
\begin{eqnarray}
K_{uu}({\bf r},{\bf r}_1)=\sum_m u_m({\bf r})u^*_m({\bf r}_1)f_1(E_m)\, ,  \nonumber \\
K_{uv}({\bf r},{\bf r}_1)=\sum_m u_m({\bf r})v^*_m({\bf
r}_1)f_1(E_m) \ . \label{def-K}
\end{eqnarray}
The kernels obey the symmetry $ K_{uu}({\bf r},{\bf
r}_1)=-K_{vv}^*({\bf r},{\bf r}_1) =K_{uu}^*({\bf r}_1,{\bf r}) $
and $K_{uv}({\bf r},{\bf r}_1)=K_{vu}^*({\bf r},{\bf
r}_1)=K_{uv}({\bf r}_1,{\bf r})$ due to the symmetry of the
functions $ u_{E}\rightarrow v^*_{-E}\ , \; v_{E}\rightarrow
-u^*_{-E} $ with respect to $E\to -E$. The excitation charge is
also expressed through $u_n$ and $v_n$, $ N=\frac{1}{2}\sum_{n}
N_n $, where
\begin{equation}
N_{n}=-\int dV  \left[ u^*_n({\bf r} )u_n({\bf r} )- v^*_n({\bf r}
)v_n({\bf r} )\right] f_1(E_n) \label{Nnn}
\end{equation}
is the number of excitations on the state $n$, it determines the
average excitation charge $Q_n=eN_{n}$ carried by the state. The
sum runs over the states with both positive and negative energies.
It can be written as a sum over only positive-energy states:
$N=\sum_{n, E_n>0} N_{n}$.

\paragraph{Andreev states. --} The normal conductor has a single
transport mode parameterized by the coordinate $x$ along it. The
eigenstates in a short low-transparency junction have the form of
narrow particle-like and hole-like resonances such that only two
of them fit into the sub-gap range of $|\Delta|$. Due to proximity
to the superconducting leads these states transform into the
Andreev states localized over the distances of the order of $\xi$
near the junction. All other resonances are essentially the same
as in the normal state; thus they can be ignored if one is
interested in energies of the order of the superconducting gap.
The two Andreev states will be labeled with $n=+1$ and $n=-1$.
One of them has a positive and other has a negative energy $E_1=-
E_{-1}>0$. In what follows we assume that the transparency is so
small that ${\cal T}\ll d/\xi$. In this limit the Andreev states
are mostly concentrated inside the normal conductor. We will also
consider energies much smaller that the superconducting gap
because it is this energy range where the supercurrent enhancement
is most pronounced. In this case the kernels Eqs. (\ref{def-K}) in
Eqs. (\ref{inteq1}) and (\ref{inteq2}) contain contributions only
from the two Andreev states, $n=\pm 1$. Due to the symmetry with
respect to $E\to -E$,
\begin{eqnarray*}
K_{uu}(x,x_1) =[u_1(x)u^*_1(x_1)-v^*_1(x)v_1(x_1)]f_1(E_1) \, , \\
K_{uv}(x,x_1) =[u_1(x)v^*_1(x_1)+v^*_1(x)u_1(x_1)]f_1(E_1)\, .
\end{eqnarray*}
Denoting $ u_1=u^+ e^{ip_u x}+u^- e^{-ip_u x}$ and $ v_1=v^+
e^{ip_v x}+v^- e^{-ip_vx} $ and neglecting the difference between
the momenta of $u$ and $v$ in the kernels, $p_u=p_v\equiv p$ (for
a short junction, $p_u-p_v\sim E/v_F \ll 1/d$) we find
\begin{eqnarray*}
&&\int_{-d/2}^{d/2} \! dx_1 \left[ K_{uu}(x,x_1)u_1(x_1) -
K_{uv}(x,x_1)v_1(x_1)\right] \\
&& \qquad \qquad=-u_1(x) N_{1}  -v_1^*(x)M_{1}\, , \\
&& \int_{-d/2}^{d/2} \! dx_1 \left[K_{uu}^*(x,x_1)v_1(x_1) +
K_{uv}^*(x,x_1)u_1(x_1)\right]\\
&&\qquad  \qquad =-v_1(x)N_{1}
+u_1^*(x) M_{1}\, ,
\end{eqnarray*}
$N_{1}$ being the number of excitations on the state $n=1$ while
$ M_{1}=2\int_{-d/2}^{d/2}  u_1(x_1)v_1(x_1)f_1(E_1)\, dx_1 $.

The BdG equations (\ref{inteq1}), (\ref{inteq2}) inside the
normal conductor take the form
\begin{eqnarray*}
&&\left( -\frac{\hbar^2}{2m}\frac{d^2}{dx^2} -\mu \right) u_1
- E_CM_{1} v_1^* =\phantom{-}E_1u_1 \, , \\
&&\left( -\frac{\hbar^2}{2m}\frac{d^2}{dx^2} -\mu \right) v_1 +
E_CM_{1} u_1^* =-E_1v_1
\end{eqnarray*}
where the effective chemical potential is
\begin{equation}
\mu =E_F-U_C+E_C N_{1} \label{mu}\, .
\end{equation}
For $u,v^* \propto e^{ip x}$ we find from here
\begin{equation}
p^2/2m -\mu =\pm \sqrt{E_1^2-E_g^2} \label{C-gap}
\end{equation}
where $ E_g =E_C|M_{1}| $ is the Coulomb gap. Using the solutions
of the BdG equations without the Coulomb interaction, one can show
that $M_1\sim {\cal T}$. The Andreev state energy is (see below)
$E_1\gtrsim (\hbar v_F/d){\cal T}$.
Therefore, the Coulomb gap $ E_g\sim E_C {\cal T}$ in Eq.
(\ref{C-gap}) can be neglected when $\gamma \ll 1$. In this limit
Eq.~(\ref{C-gap}) reduces to its usual form, $ p^2/2m -\mu =\pm
E_1 $, while the functions $u$ and $v$ inside the normal conductor
are decoupled.

The BdG equations
for the entire system can thus be written in a standard form
\begin{eqnarray}
&&\! \! \left[\hat H_0 -h({\bf r})E_CN_{1} \right] u_1({\bf r}) +\Delta
({\bf r}) v_1({\bf r})
 =E_1 u_1({\bf r}) , \quad \label{bdg-1}\\
&&\! \! \! \! \!\! \! \! -\left[\hat H_0 -h({\bf r})E_CN_{1}
\right] v_1({\bf r}) +\Delta^* ({\bf r}) u_1({\bf r}) =E_1v_1({\bf
r}). \quad \label{bdg-2}
\end{eqnarray}
The BdG equations (\ref{bdg-1}), (\ref{bdg-2}) determine the
energy of the Andreev bound states as a function of the phase
difference $\phi =\chi_R-\chi_L$. For low transparency, ${\cal
T}\ll d/\xi$, the energies $E_{\pm 1}=\pm (\hbar v_F/d)\epsilon_A$
in the range $E\ll |\Delta|$ have the form
\cite{KMV06,Sadovskii07}
\begin{equation}
\epsilon_A=\sqrt{z^2+{\cal T}^2\cos^2(\phi/2)} \label{eq-epsilon}
\end{equation}
where ${\cal T}={\cal T}_1/2$ is the transparency of two contacts
in series. The quantity $z$ measures the shift of the chemical
potential $\mu$ inside the conductor, Eq. (\ref{mu}), from one of
the resonances. It is defined through the equation
\begin{equation}
p_Fd/\hbar +\eta +(d/\hbar v_F)[E_CN_1-U_C] =\pi k +z \label{p0}
\end{equation}
where $\eta =\arctan\sqrt{{\cal T}_1}$ is the scattering phase
shift and $k$ is an integer. The range of applicability of
Eq.~(\ref{eq-epsilon}) is $\epsilon_A \ll d/\xi$.

As it follows from  Eq. (\ref{Nnn}), $U_C=U_C^\prime +2E_CN_1 $ where
$U_C^\prime =2E_C(N^\prime -N_0) $ and $ N^\prime =\sum_{n \ne 1}
N_{n} $. Let us denote
\[
y=p_Fd/\hbar +\eta -dU_C^\prime/\hbar v_F -\pi k \, .
\]
The quantity $y$ depends on the gate voltage through the
background charge $N_0$. The excitation number $N^\prime$ changes
little if the chemical potential varies within one resonance,
$z\lesssim 1$. Therefore, the chemical potential shift $y$ can
 be considered as a parameter controlled by the gate voltage.

For a localized state \cite{KMV06}
\begin{equation}
N_{1}=\left(\frac{\partial E_1}{\partial \mu }\right)_{\!\!\phi}
f_1(E_1) \, .\label{charge1}
\end{equation}
The subscript $\phi$ means that the phase is kept constant. As it
follows from Eq.~(\ref{mu}) for the effective chemical potential
of the Andreev state, $ \partial \mu/\partial z =\hbar v_F/d$.
Therefore, Eq.~(\ref{p0}) yields
\begin{equation}
y =z+\gamma \frac{\partial \epsilon_A}{\partial z}f_1(E_1)\, .
\label{def-z}
\end{equation}
We consider zero temperatures and put $f_1(E_1)=1$. This can be
done if temperature satisfies $T\ll {\cal T}\delta E\sim \Delta
{\cal T}(\xi/d)$. Equation (\ref{def-z}) is the central result of
this paper. Equations (\ref{eq-epsilon}) and (\ref{def-z})
determine $\epsilon_A$ as a function of $y$ and
Eq.~(\ref{charge1}) then determines the charge $N_{1}$ on the
Andreev level.


\paragraph{Results. --}
Equations (\ref{eq-epsilon}) and (\ref{def-z}) yield
\begin{equation}
y=(1+\gamma /\epsilon_A)\sqrt{\epsilon_A ^2-{\cal
T}^2\cos^2(\phi/2)}  \, . \label{energy1}
\end{equation}
We see that at $\gamma \ne 0$ ( i.e., due to the Coulomb
interaction) there appears a substantial region of gate voltages,
$y\lesssim \gamma$, where energy is close to its smallest possible
value, ${\cal T}\cos(\phi/2)$. This is more pronounced if the
Coulomb energy is larger than the gap $E_C/|\Delta | \gg 1$
(estimates will be given later). Indeed, since $\epsilon_A\lesssim
d/\xi$ the ratio $\gamma /\epsilon_A>E_C/|\Delta|$. Yet more
drastic effect is expected if ${\cal T}\ll \gamma$. In this case
the term $\gamma/\epsilon_A$ in brackets in Eq. (\ref{energy1})
dominates. As a result, for $y< \gamma$
\begin{equation}
\epsilon_A =\frac{{\cal T}\cos(\phi/2)}{\sqrt{1-y^2/\gamma^2}}\, .
\label{energy2}
\end{equation}
According to Eqs.\ (\ref{energy1}) or (\ref{energy2}),
$\epsilon_A$ and $z$ remain small $\epsilon_A, z \ll \gamma$ in
the range of gate voltages $y\lesssim \gamma$. Thus the energy
satisfies $E_1\ll E_C,|\Delta|$, and the Kondo regime is not
realized.

\begin{figure}
\centerline{\includegraphics[width=\linewidth]{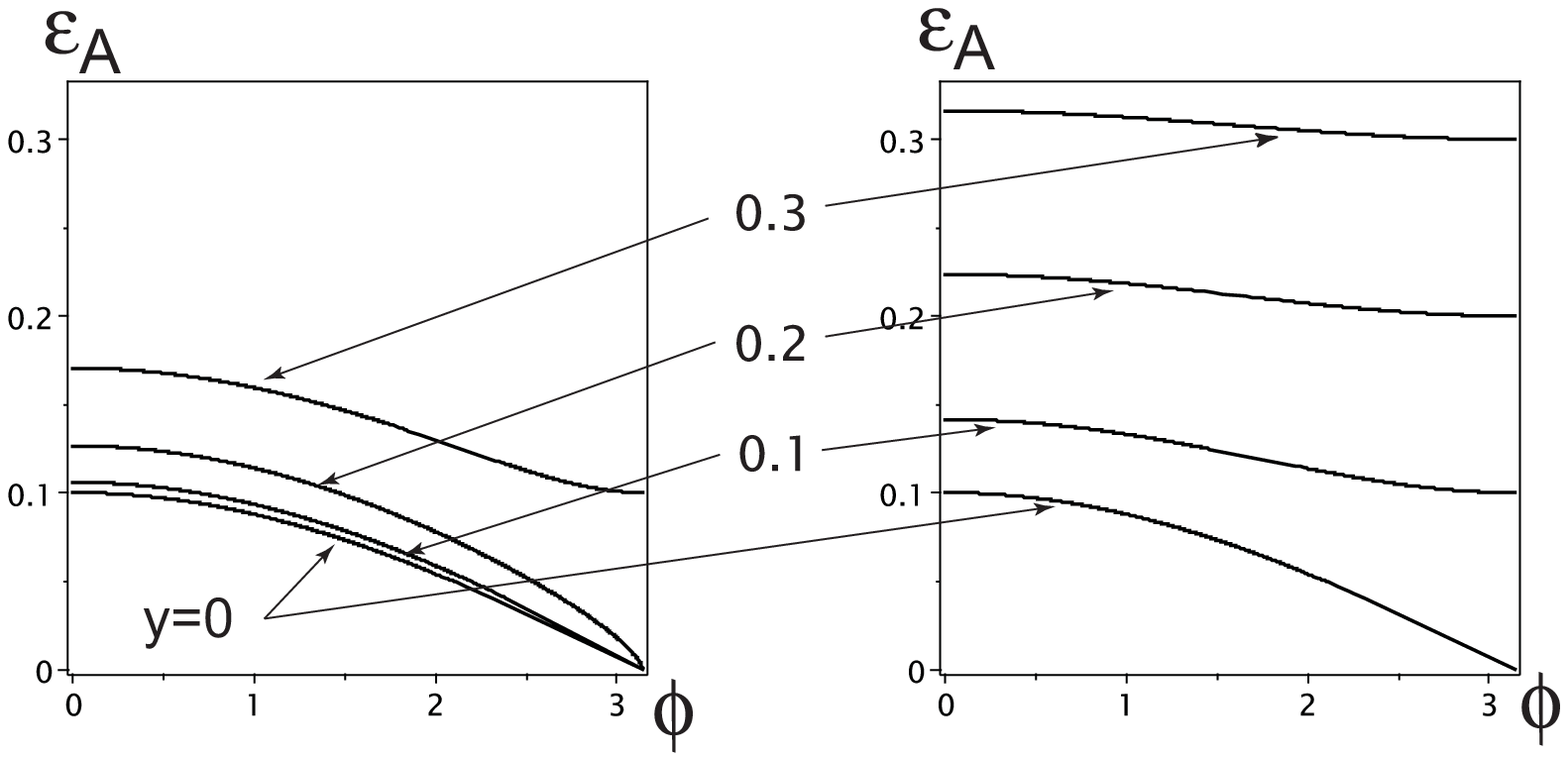} }
\centerline{\includegraphics[width=\linewidth]{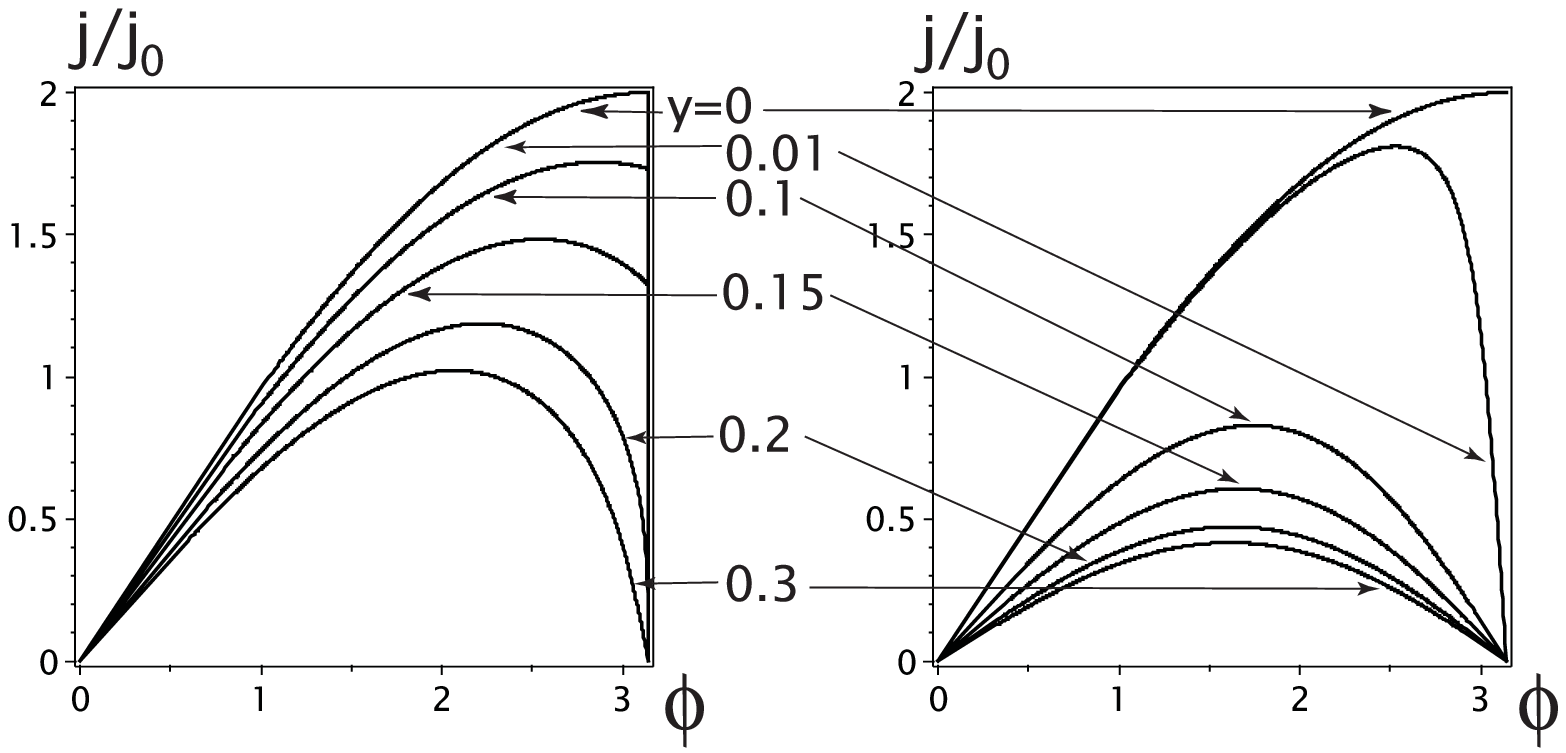} }

\caption{Andreev state energy (upper panels) and the supercurrent
(lower panels) in the presence of Coulomb interaction,
$\gamma=0.2$ (left panels) and without it, $\gamma =0$ (right
panels) as  functions of the phase difference $\phi$ and the gate
voltage $y$ for $\mathcal{T}=0.1$. The curves for $\gamma =0$ and
$\gamma =0$  coincide at $y \to 0$. The curves for the
supercurrent at $\gamma =0.2$ for  $y=0$ and $y=0.01$ overlap,
while at $\gamma =0$ they significantly differ at $\phi$ close to
$\pi$. \label{fig1}}

\end{figure}

Figure~\ref{fig1} (upper panels) shows the solutions of
Eqs.~(\ref{energy1}) for $\mathcal{T}=0.1$ and for two values of
the Coulomb interaction, $\gamma =0.2$ (left panel) and $\gamma
=0$ (right panel). One observes that $\epsilon_A\to 0$ as $\phi$
approaches $\pi$ if $y<\gamma$. The decrease in energy is, of
course, cut off when $|\phi -\pi|\sim (T/|\Delta|)(d/\xi {\cal
T})$ because of the decreasing factor $f_1(E_1)$ in Eq.
(\ref{def-z}). This energy behavior has a crucial impact on the
supercurrent given by the expression \cite{Beenakker91}:
\begin{equation}
j_s=-\frac{2e}{\hbar} \left(\frac{\partial E_1}{\partial
\phi}\right)_{\!\!\mu}f_1(E_1) =j_0\frac{\mathcal{T} \sin
\phi}{\epsilon_A} \, ,  \label{supercurrent1}
\end{equation}
where $j_0 =ev_F{\cal T}/2d$. Since the energy remains small in a
wide range of gate voltages, the supercurrent is essentially
enhanced as compared to its zero-Coulomb value for the same gate
voltage. In particular, for comparatively large Coulomb
interaction, $\gamma \gg {\cal T}$, we have from
 Eq.~(\ref{energy2})
\[
j/j_0 =2\sqrt{1-y^2/\gamma^2}\, \sin (\phi/2)\, .
\]
The critical current $j_c=2j_0\sqrt{1-y^2/\gamma^2}$ is reached at
$\phi =\pi$. It is roughly by factor $\gamma/{\cal T}$ larger than
its value without the Coulomb interaction, $j_c= j_0{\cal T}/y$,
reached at $\phi =\pi/2$ for the same gate voltage $y\gg {\cal
T}$. The supercurrent Eq.~(\ref{supercurrent1}) is shown in Fig.
\ref{fig1} (lower panels) for $\mathcal{T}=0.1$ and for two value
of the Coulomb interaction, $\gamma =0.2$ (left panel) and $\gamma
=0$ (right panel). One observes significant enhancement of the
supercurrent for a given gate voltage.

\paragraph{Discussion. --}
We have shown that charging of the Andreev level leads to a strong
enhancement of the supercurrent in a wide range of gate voltages
(parameterized by the quantity $y$). The charge of the Andreev
state compensates the deviation of the chemical potential caused
by the change in the gate voltage thus preserving the resonance
transmission of the double barrier structure and increasing the
transparency of the device.

The Coulomb interaction is characterized by the dimensionless
parameter $\gamma =dE_C/\hbar v_F$. The upper estimate gives
$\gamma \sim e^2/\hbar v_F \lesssim 1$. Our approximation requires
small $\gamma$. Nevertheless, the factor $ \gamma/{\cal T} $ that
determines the enhancement of the supercurrent can reach quite
large values due to low transparency. A substantial enhancement of
supercurrent occurs already when  $E_C\gtrsim |\Delta|$. A typical
gap $\Delta$ (as for Al) is about 1 K. The Coulomb energy
$E_C=e^2/2C$ of 1 K corresponds to the capacitance $ C \approx
8\cdot 10^{-4}$ cm. Therefore, to have $E_C\gg \Delta$ the
characteristic size of the normal conductor should be smaller than
$\sim 10^{-3}$~cm which can be easily realized in practice.

\acknowledgments

We thank I. Sadovskyy, V.~Shumeiko, and A.~Zazunov for stimulating
discussions. This work was supported by the Russian Foundation for
Basic Research under grant 09-02-00573-a, by the Program ``Quantum
Physics of Condensed Matter'' of the Russian Academy of Sciences,
by the Academy of Finland Centers of Excellence Program, by the
U.S. Department of Energy Office of Science under Contract No.
DE-AC02-06CH11357, and by Norwegian Research Council through the
program on sensors and detectors.

\end{document}